# ELECTROMAGNETIC RADIATION BY ELECTRONS IN THE CORRUGATED GRAPHENE

**S.A. Ktitorov[1,2], R.I. Mukhamadiarov[1]**
[1]Ioffe Institute, St Petersburg, Russia,
[2]St. Petersburg Electrotechnical University LETI, St. Petersburg, Russia
ktitorov@mail.ioffe.ru

The electromagnetic radiation of electrons in the corrugated graphene in the presence of the transport electric current in the ballistic regime is studied. Radiation of the similar nature can be observed in undulator and wiggler. We considered here an impact of the ripples in the monolayer graphene on its electromagnetic properties. The electromagnetic radiation was actually calculated with a use of the standard electromagnetic theory. Two cases, of regular and random structures are analyzed. Nonlinear relation between the random height function $h(x,y)$ and the gauge field is shown to create the radiation frequency distribution central peak. Few mechanisms of ripples formation in monolayer graphene were considered. The ripples are considered as an incommensurate superstructure in the two-dimensional crystal, appearing as a result of forming of periodic solutions in the in-plane optical phonon subsystem. Possible instability of the flexural subsystem is discussed as well.

## 1. Introduction

The recent experimental study has shown that a flat geometry of graphene is unstable that leads to forming of corrugations: topological defects and ripples [1]. Graphene can be viewed as a crystalline membrane. An ideal flat 2D-crystal could not exist at a finite temperature [1], and the long range order in the graphene takes place because of ripples and topological defects. Both of these factors facilitate to achieve thermodynamic stability in graphene [2] and they could be induced to relief the strain of the membrane. As far as we know, no widely accepted model of ripples in monolayer graphene exists. Crumpled membrane theory does not work for graphene at actual temperatures. Some authors try to obtain a ripple-like solution from the renormalization group approach, but the only result is appearance of some characteristic length of a correct magnitude. The phenomenon under consideration looks too simple to be described with a use of the sophisticated fluctuation theory: we guess that the ripple theory has to be the mean field one. Rather similar problem of the undulations in biological membranes is solved by taking account of the internal degrees of freedom [3]. One of such fluctuating degrees of freedom was the membrane thickness. In the case of graphene a natural candidate on this position is the in-plane transverse optical phonon. The spatial period of the ripple is determined by the ratio of the coefficients in the thermodynamic potential derivative expansion. Out-plane subsystem acquires a periodicity due to the interaction between the subsystems.

Despite of the "quasirelativistic" character of the spectrum, the ratio of Fermi velocity to the speed of light is much smaller, than unity and we can neglect a retardation of the electromagnetic radiation. This gives us a reason to consider a motion of electrons in graphene within the classical approach. The most important mechanisms are bremsstrahlung, cyclotron, and undulator radiation [4]. The emission mechanism under the consideration resembles one in the undulator [5] but practically without retardation. We analyzed radiation using a few distinct models. At first model we consider geometric mechanism, directly connected with the presence of undulations, at the second – the pseudo gauge field effect making trajectory to get curved in the base plane. Both regular and random ripple structures are considered. In the next sections we will consider various models of electrons motion in corrugated graphene and will derive formula for the radiation intensity.

## 2. Electrodynamics of radiation in graphene
## 2.1. Geometric mechanism

Anomalously high electrons mobility in graphene [4] leads to the mean free path of micron value [6]. For the mean ripples period of 50 nanometers, the ratio of the free path to period of the structure is about 20, which leads to the 10% spectral line broadening. The ballistic regime is implemented for the graphene sample of about several microns. Electrons motion through the rippled graphene sheet induces an electromagnetic radiation in the terahertz range [4,7,8]. The mechanism of formation of the bremsstrahlung radiation in graphene is similar to one in the undulator or wiggler [9]. While the electron trajectory in an undulator and wiggler deviates from the direct line due to the periodic system of the dipole magnets, electron trajectory in graphene is getting curved due to the ripples. A spatial period of ripples in graphene could reach several hundreds of nanometers [1]. This makes the semiclassical approach feasible.

The Fermi velocity vector in the vicinity of the Dirac points in graphene has the constant absolute value of $10^8$ cm/s. However, its orientation changes with time leading to a time dependence of the vector components that is responsible for emission of the electromagnetic waves.

The ripple average amplitude is about 1 nm [10], whereas period $L \sim 50$ nm. Taking this into account, we could assume that velocity preserves a constant value in the direction of applied field. In other words, electrons in graphene could be considered as an oscillator with the Fermi velocity. The real graphene sample has random corrugations period that deviates a little from mean value. The chaotic surface could be considered as a superposition of sinusoids with their own period and height. This approach widely used in the wave analysis in radiophysics. In order to describe the random process of electromagnetic radiation with limited spectrum we introduce the random function $h(x)$ that plays a role of the ripples height relative to the base plane. Assuming inhomogeneities to be one-dimensional and taking into account small ratio of $\frac{h_m}{L} \ll 1$ we can write the following equations for the velocity

$$v_x = v_F \quad v_z = v_F \frac{d}{dx} h(x(t)). \tag{1}$$

This means that the electromagnetic radiation propagates in normal direction and infinitesimal changes in in-plane velocity components give the effects of higher order [11] and can be neglected.
We calculate the magnetic field induced by moving charge at the deliberately chosen point with use of retarded potentials. The Fourier transform of the vector potential reads:

$$\mathbf{A}_\omega = \frac{e\,e^{ikr}}{cr} \int_0^\infty \mathbf{v}(t) e^{i(\omega t - \mathbf{k}\mathbf{r}_0)} dt. \tag{2}$$

where $c$ – speed of light, $e$ – electrons charge, $\mathbf{r}$ – radius vector to a view point, $\mathbf{r}_0$ – to a moving charge.
Taking (1) into account and excluding the retardation effect we have:

$$A_z = \frac{e\,e^{ikr}}{cr} v_F \int_0^\infty \frac{d}{dx} h(x(t)) e^{it\omega} dt. \tag{3}$$

The other two components will be either constant or equals zero and do not contribute in generation of electromagnetic waves. The Fourier component of the magnetic field becomes:

$$B_y = -ik_x \frac{e\, e^{ikr}}{cr} v_F \int_0^\infty \frac{d}{dx} h(x(t)) e^{it\omega} dt \qquad (4)$$

Magnetic field enters at the intensity formula in squared form:

$$|B_y|^2 = \left(k_x \frac{e}{cr} v_F\right)^2 e^{ikr} e^{-ikr} \int_0^\infty dt\, \frac{d}{dx} h(x(t)) e^{i\omega t} \cdot \int_0^\infty dt'\, \frac{d}{dx'} h(x'(t')) e^{-i\omega t'} \qquad (5)$$

In case of random surface of the graphene, the squared absolute value can be represented as a multiplication of two equations where distribution of random function $h(x)$ could be considered as realizations, which could differ even in the same point [12]. Therefore, while calculating the field intensity, the relation between them could be represented in terms of the correlation function. We introduce it by averaging (5) over the ripples configuration:

$$\langle |B_x|^2 \rangle = -k_y^2 \frac{e^2 v_F^2}{c^2 r^2} \int_0^\infty dt \int_0^\infty dt'\, \frac{d^2}{dx^2} \langle h(x(t)) h(x'(t')) \rangle e^{i\omega(t-t')}, \qquad (6)$$

$$\langle |B_y|^2 \rangle = -k_x^2 \frac{e^2 v_F^2}{c^2 r^2} \int_0^\infty dt \int_0^\infty dt'\, \frac{d^2}{dx^2} \langle h(x(t)) h(x'(t')) \rangle e^{i\omega(t-t')}, \qquad (7)$$

where the angle brackets stand for configuration averaging. The random process $h(x)$ could be considered as Gaussian and stationary with the correlator

$$\langle h(x) h(x') \rangle = K(x - x') = K(\xi). \qquad (8)$$

where correlation function value depends on relative coordinate $x - x'$, which means that statistical characteristics are invariant under a shift along the *OX*. Assuming this process to be one with the narrow band spectrum we write the correlator in the form:

$$K(\xi) = \langle h^2 \rangle e^{-\alpha|\xi|} \cos \gamma \xi, \qquad (9)$$

where $\langle h^2 \rangle$ – dispersion, $\alpha$ is the inverse correlation radius and $\gamma$ – the inverse mean period and $\alpha \ll \gamma$. Making the substitution $\xi = x - x'$, $\tau = t - t'$, $T = (t+t')/2$ we can rewrite formula (6) in the form:

$$\langle |B_x|^2 \rangle = \left(\frac{k_y^2 e^2}{r^2}\right) \frac{v_F^2}{c^2} \frac{\omega^2}{v_F^2} \langle h^2 \rangle \int_0^\infty dT \int_0^\infty d\tau\, e^{-\alpha v_F |\tau|} \cos(\gamma v_F \tau)\, e^{iT(\omega-\omega')} e^{i\frac{\tau}{2}(\omega+\omega')}. \qquad (10)$$

Integration over difference variable makes a Fourier transform whereas integration over sum variable is simply the time requires for electron to travel through graphene sample:

$$\langle |B_x|^2 \rangle = -\delta(\omega-\omega') \frac{k_y^2 e^2 \langle h^2 \rangle \omega^2}{c^2 r^2} \left( \frac{\alpha v_F}{\alpha^2 v_F^2 + (\omega + \gamma v_F)^2} + \frac{\alpha v_F}{\alpha^2 v_F^2 + (\omega - \gamma v_F)^2} \right). \qquad (11)$$

Spectral distribution of electromagnetic radiation

$$\frac{d^2\mathcal{E}}{d\omega do} = \frac{c}{4\pi^2}\left\langle |B_x|^2 \right\rangle r^2 \qquad (12)$$

Substituting (11) in (12) we obtain the formula for the spectral intensity of electromagnetic radiation in rippled graphene:

$$\frac{d^2\mathcal{E}}{d\omega do} = \delta(\omega - \omega')\frac{k_y^2 e^2 \left\langle h^2 \right\rangle \omega^2}{4\pi^2 c}\left(\frac{\alpha v_F}{\alpha^2 v_F^2 + (\omega + \gamma v_F)^2} + \frac{\alpha v_F}{\alpha^2 v_F^2 + (\omega - \gamma v_F)^2}\right) \qquad (13)$$

Integrating (13) by frequencies and angle

$$P = \frac{c}{4\pi}\int_{-\infty}^{\infty}d\omega'\int_{-\infty}^{\infty}d\omega\int_{0}^{2\pi}d\Omega \left\langle |B_x|^2 \right\rangle r^2, \qquad (14)$$

we obtain the relation for power of electromagnetic radiation for one electron:

$$P_e = \int_{-\infty}^{\infty}d\omega\int_{0}^{2\pi}d\Omega\cos\theta \frac{e^2\omega^4 \left\langle h^2 \right\rangle}{4\pi c^3}\left(\frac{\alpha v_F}{\alpha^2 v_F^2 + (\omega + \gamma v_F)^2} + \frac{\alpha v_F}{\alpha^2 v_F^2 + (\omega - \gamma v_F)^2}\right). \qquad (15)$$

We see that the most part of electromagnetic radiation comes with frequency corresponding to the mean period of the ripples structure with some broadening due to its random character. The term in brackets yields the Lorentz distribution.

In the limit of $\alpha \to 0$ the chaotic surface becomes a regular sine and we get the formula

$$P_e = \int_{-\infty}^{\infty}d\omega\int_{0}^{2\pi}d\Omega\cos\theta \frac{e^2\omega^4 h_m^2}{4c^3}\delta(\omega - \gamma v_F) = \frac{\pi e^2(\gamma v_F)^4 h_m^2}{c^3} \qquad (16)$$

The obtained formulae establish a relation between the radiation spectral density $\frac{dP}{d\omega}$ and the correlator of ripples. This can be used for investigation of the ripples morphology.

For ripples period $L = 50$ nm, maximum amplitude $h_m = 1$ nm and area of graphene membrane $S = 10^{-4} cm^2$, we obtain power of electromagnetic radiation of order several mW.

### 2.1. Gauge mechanism

In this part of our work we consider the synthetic gauge field effect on the electromagnetic waves emission. The nature of these fields is following. Electronic states in flat graphene can be written by means of the tight binding model equations, which due to linear spectra in Dirac point vicinity take form of Dirac-Weyl equations. The graphene membrane bending modifies the electronic states spectra. Moreover, this influence can be represented by introducing vector potential, rotor of which people called synthetic magnetic field [1]. In fact, these "fields" create new energy levels, similar to Landau levels, and mimics Ahoronov-Bohm effect [1]. The elasticity theory gives the following relations between the out-of-plane displacement $h(y,z)$ and the gauge field vector potential components[13]:

$$A_y = -\frac{\beta}{a}\frac{dh}{dx}\frac{dh}{dy}, \quad A_x = \frac{1}{2}\frac{\beta}{a}\left(\left(\frac{dh}{dx}\right)^2 - \left(\frac{dh}{dy}\right)^2\right), \qquad (17)$$

$$\mathcal{H}_z = \frac{\partial A_y}{\partial x} - \frac{\partial A_x}{\partial y}.$$

where $a$ is the lattice constant, $\beta$ is a dimensionless parameter. It now becomes apparent that the gauge fields in graphene take place only when structure is inhomogeneous in both in-plane directions. In other words graphene sheet has to have "humps and hollows" structure.

Here we consider the case of regular undulation in both directions:

$$h(x,y) = h_m \sin \gamma x \sin \gamma y. \tag{18}$$

Substitution of $h(x,y)$ from (18) into (17) gives an expression for the synthetic magnetic field:

$$\mathcal{H}_z = \mathcal{H}_0 \sin(2y\gamma)(-2 + \cos(2x\gamma)), \tag{19}$$

where the amplitude is determined as follows:

$$\mathcal{H}_0 = \frac{h_m^2 \gamma^3}{2a}. \tag{20}$$

The synthetic field $\mathcal{H}$ can be expressed in terms of the "real" field $H$ according to the relation

$$H = \frac{c\hbar}{e} \mathcal{H} \tag{21}$$

After this link has been made, we can modify the couple of equations from [14] and define undulator-like trajectory for "massless" electrons in graphene

$$\begin{aligned} \frac{\mathcal{E}}{v_F^2} \frac{d^2 x}{dt^2} &= -e \frac{dy}{dt} H \cos \gamma x, \\ \frac{\mathcal{E}}{v_F^2} \frac{d^2 y}{dt^2} &= e \frac{dx}{dt} H \cos \gamma x, \end{aligned} \tag{22}$$

where $\mathcal{E}$ is the electron energy. The solution reads:

$$\cos \gamma x = \cosh \gamma y - \frac{1}{k} \sinh \gamma y. \tag{23}$$

The resulted trajectory slightly differs from simple sine. Making an assumption that $y\gamma \ll 1$ we can simplify the last formula:

$$y = y_m \sin^2 \frac{\gamma x}{2}, \tag{24}$$

where the amplitude of deviations is

$$y_m = \frac{2k}{\gamma} = \frac{2 v_F \hbar h_m^2 \gamma}{a \mathcal{E}}. \tag{25}$$

The estimations yield maximum amplitude of in-plane deviations to be much smaller then mean period of the ripples and have values close to ripples amplitude. This makes us think that electrons actually move along complex helix-like trajectory.

These formulae show that gauge fields in graphene have real influence on electrons trajectory and induce in-plane oscillations. The magnitude of the effect depends from sample parameters and free electrons energy.

Now, with (22) we are able to derive formula for power. Substituting the expression for the oscillating electron velocity $v_y = v_F y'$ into (17) we obtain the formula for the radiation vector potential

$$A_y = \frac{e\, e^{ikr}}{cr} \int (y_m \gamma) v_F \sin(\gamma v_F t) e^{i\omega t}. \qquad (26)$$

Thus, we have an expression for the radiation power spectral density for the case of the harmonic ripples structure

$$P = \frac{1}{4\pi} \int d\omega \int d\theta \frac{e^2 v_F^2 \omega^2}{2c^3} (y_m^2 \gamma^2) \cos\theta \left( \delta(\omega + \gamma v_F) + \delta(\omega - \gamma v_F) \right), \qquad (27)$$

Finally, integrating the expression and substituting (25) for $y_m$, we derive a formula for the electromagnetic radiation power in terms of the sample parameters

$$P = 2\pi^2 \frac{e^2 v_F^4}{c^3} \frac{h_m^2}{L^4} \left( \left( \frac{v_F \hbar / a}{\mathcal{E}} \right) (h_m \gamma) \right)^2 \qquad (28)$$

To understand the role of gauge fields in whole picture of the radiation process, it will be useful to compare power formula obtained in geometric model with (28)

$$\frac{P_{cal}}{P_{geom}} = \frac{1}{8\pi^3} \left( \frac{v_F \hbar}{a\mathcal{E}} \frac{h_m}{L} \right)^2. \qquad (29)$$

This ratio would be of order of unity for the same parameters of graphene structure at $\mathcal{E} =$ 0.1 eV energies.

Calculations for the case of the random function $h(x.y)$ are carried out similarly to the case of the geometric mechanism but with one important distinction: a nonlinear relation (17) between the narrow band spectrum random function $h(x.y)$ and the gauge field $A$ induces appearance of central peak of radiation. Really, let us determine the quadratic relation between the random functions:

$$g(t) = f(h(t)). \qquad (30)$$

When $g(h) = bh^2$, the correlator $<g(t)g(0)>$ can be expressed as follows [12]:

$$\langle g(t) g(0) \rangle = b^2 \sigma^4 \left[ 1 + 2 \langle h(t) h(0) \rangle^2 \right], \qquad (31)$$

where $\sigma$ is the $h(t)$ process variance $\sigma = \langle h^2 \rangle$ and $b$ is some constant.

We do not aware about existing experimental data or theoretical results regarding actual correlation function's $\langle h(x) h(x') \rangle$ form. All we can do is estimate dispersion and correlation radius. A reasonable choice would be the simplest form that does not lead to any disagreement. We choose the exponential decay for geometry model, i.e. Lorentz-shaped energy spectra. In case of gauge fields, the exponential correlator yields singularity: it corresponds to infinitely large gradients of random function[15]. Thus, we choose the Gaussian correlator form:

$$K(x) = \frac{\langle h^2 \rangle \alpha^3}{8\pi\sqrt{\pi}} \exp\left(-\frac{x^2 \alpha^2}{4}\right) \cos(\gamma x) \qquad (32)$$

Substituting (32) in (31) we have

$$\langle g(t)g(0)\rangle = b^2 \sigma^4 \left[1 + 2\left(\frac{\langle h^2 \rangle \alpha^3}{8\pi\sqrt{\pi}}\right)^2 \exp\left(-\frac{2(v_F t)^2 \alpha^2}{4}\right) \cos^2(\gamma v_F t)\right] \qquad (33)$$

Taking into account formulae (17) we obtain for gauge field random process:

$$\int_{-\infty}^{\infty} dx\, e^{ikx} K^2(x) = b^2 \sigma^4 \left(\frac{\langle h^2 \rangle \alpha^3}{8\pi\sqrt{\pi}}\right)^2 \int_0^{\infty} dx\, e^{-\left(\frac{\alpha^2 x^2}{2}\right)} \left[\cos(kx) + \left[\cos(k+2\gamma)x + \cos(k-2\gamma)x\right]\right] \qquad (34)$$

Carrying out the integration we obtain the spectrum comprising the central and high-frequency components:

$$P \propto P_0 + b^2 \sigma^4 \frac{\langle h^2 \rangle^2 \alpha^5 k^3}{16\pi^2 \sqrt{\pi}} \left[e^{-\frac{k^2}{\alpha^2}} + \left[e^{-\frac{(k+2\gamma)^2}{\alpha^2}} + e^{-\frac{(k-2\gamma)^2}{\alpha^2}}\right]\right]. \qquad (35)$$

The pre-exponent term $k^3$ in (35) accounts for differentiation in (17) formula. On the fig. 1 we presented the spectral distribution of a radiation power, divided by $k^3$.

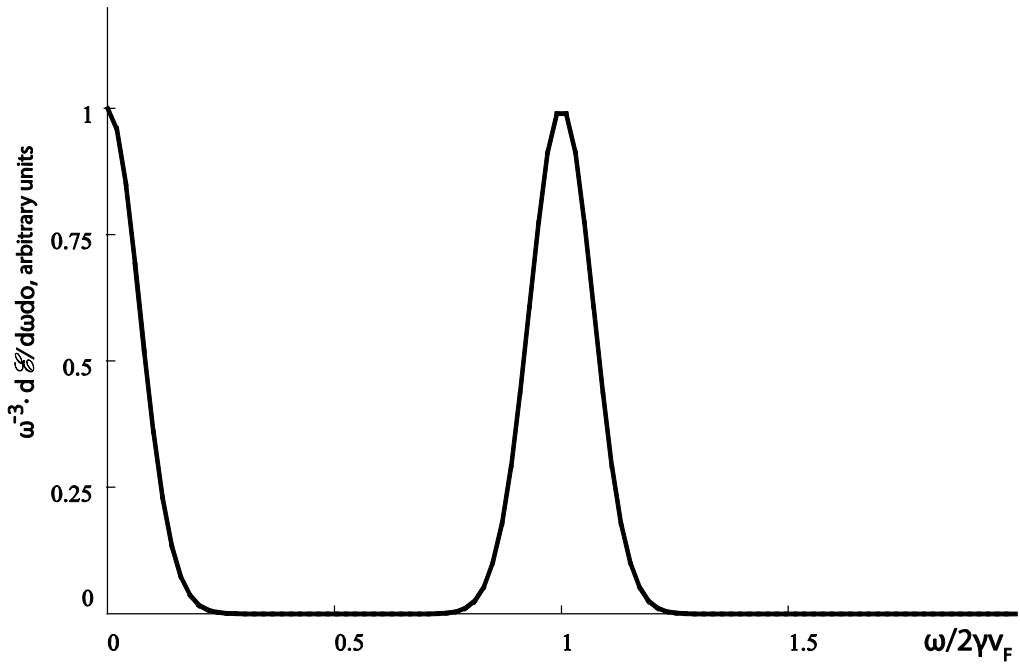

**Fig. 1**. Central and side radiation peaks in gauge model

Thus, the quadratic relation between velocity and the Monge variable leads to a rise of the electromagnetic radiation central peak.

## 3. Ripples model
### 3.1. In-plane phonon instability

Equilibrium state of the membrane is determined by the effective potential in the Monge representation [3]:

$$F[h] = \int d^2x \left[ \frac{\kappa}{2} (\nabla^2 h)^2 + \frac{K_0}{8} \left[ P_{\alpha\beta} (\partial_\alpha h \cdot \partial_\beta h) \right]^2 \right], \tag{36}$$

where $\kappa$ is the bending rigidity, $h$ is the membrane height relative to the base plane (Monge variable), and the second term is Gaussian curvature.

$$P_{\alpha\beta} = \delta_{\alpha\beta} - q_\alpha q_\beta / q^2, \quad K_0 = 2\mu(2\mu + D\lambda)/(2\mu + \lambda), \tag{37}$$

$P_{\alpha\beta}$ is - normal projector, $\mu$ and $\lambda$ are the Lame elastic modules. Here we assume this subsystem to be stable being isolated. In our case the natural candidate on the symmetry violating system position is the in-plane transverse optical phonon. The simplest Landau functional containing a certain spatial scale (apart from the lattice spacing) in a crystal without the Lifshitz invariant is one suggested for incommensurate ferroelectrics [16]:

$$F_{incomm} = \int dxdy \left[ \frac{a_2}{2} \phi^2 + \frac{a_4}{4} \phi^4 + \frac{c_2}{2} (\nabla \phi)^2 + \frac{d_2}{2} (\nabla^2 \phi)^2 \right]. \tag{38}$$

Here $a_2 < 0$, $a_4 > 0$, $c > 0$, $d > 0$. Varying this functional we obtain the equilibrium equation:

$$d_2 \frac{d^4 \phi}{dx^4} + c_2 \frac{d^2 \phi}{dx^2} + |a_2|\phi - a_4 \phi^3 = 0. \tag{39}$$

We consider here one-dimensional solutions. For the chosen signature of the coefficients there exists a periodic solution with the characteristic period value of order $\sqrt{d_2/c_2}$. This violation of the translation symmetry can be transferred to the out-plane subsystem by means of the interaction. Simplest interactions that do not break the spatial inversion symmetry read:

$$g \int dxdy \phi \Delta_2 h \tag{40}$$

and

$$G \int dxdy \phi^2 (\Delta_2^2 h)^2. \tag{41}$$

The interaction (40) is more effective, but is not easy to derive. Now we consider another scenario.

### 3.2. Out-plane phonon instability

In corresponding with the Landau-Peierls-Mermin-Wagner theorem, the logarithmic divergencies indicating instability of the flat state are characteristic for two-dimensional systems. In particular, the bending rigidity constant will be renormalized [3]:

$$\kappa_r = \kappa_0 - \frac{3k_B T}{4\pi} \ln L/a. \tag{42}$$

where $a$ is the lattice spacing, $L$ is the sample size. We believe that fluctuations can make absolute value $\kappa$ negative. This may induce instability, which stabilizes by non-linear terms as in phase transitions theory.

## Conclusion

We considered here an impact of the ripples in the monolayer graphene on its electromagnetic properties. Two mechanisms of the undulator-like radiation are considered:

geometric mechanism, directly connected with the presence of undulations and the pseudo gauge field effect making trajectory to get curved in the base plane. The electromagnetic radiation was actually calculated with a use of the standard retarded potential. For both of mechanisms two cases, of regular and random structures are analyzed. Nonlinear relation between the random height function *h(x,y)* and the gauge field A is shown to create the radiation frequency distribution central peak. The two models of ripples origination were proposed. Our results can be used for study of ripples morphology and for generation of terahertz radiation.